\documentclass[11pt,twocolumn,twoside]{osajnl}

\journal{josaa} 

\setboolean{shortarticle}{false}

\usepackage{color}
\usepackage{graphicx}
\usepackage{soul}
\usepackage{amsmath}
\usepackage{enumerate}
\usepackage{cancel}
\usepackage{physics}

\def\be{\begin{equation}}
\def\te{\end{equation}}
\def\ee{\end{equation}}
\def\ba{\begin{eqnarray}}
\def\bea{\begin{eqnarray}}
\def\nn{\nonumber\\}
\def\tea{\end{eqnarray}}
\def\ea{\end{eqnarray}}
\def\eea{\end{eqnarray}}
\def\vecx{\vec{x}}
\def\vecp{\vec{p}}
\def\vecq{\vec{q}}

\def\mbp{\mathbf{p}}

\def\mbk{\mathbf{k}}

\def\vpp{\varphi_{\vec{p}}}

\def\EPpq{\varepsilon_{\vec{p}-\vec{q}}}
\def\EPpk{\varepsilon_{\vec{p}-\vec{k}}}

\def\EPqki{\varepsilon_{\vec{q}-\vec{k}}}
\def\UPa{\uparrow}
\def\DWa{\downarrow}

\title{A coherent full microwave scattering formulation for random layered media}

\author[1,2]{Esteban Calzetta}
\author[3*]{Mariano Franco}

\affil[1]{Universidad de Buenos Aires, Facultad de Ciencias Exactas y Naturales, Departamento de F\'isica. Ciudad Universitaria, Int. G\"uiraldes 2160, Buenos Aires, CABA, C1428EGA, Argentina}
\affil[2]{CONICET-Instituto de F\'isica de Buenos Aires (IFIBA). Buenos Aires, Argentina.}
\affil[3]{CONICET-Universidad de Buenos Aires, Instituto de Astronomía y Física del Espacio (IAFE), Ciudad Universitaria, Av. Cantilo S/N, Buenos Aires, C1428ZAA, Argentina.}

\affil[*]{Corresponding author: mfranco@iafe.uba.ar}

\begin{abstract}
We present a fully coherent, analytic model of the backscattering intensity in all HH, HV, VH and VV channels, for the volume scattering of radiation from a layer of finite thickness, such as a vegetation layer over bare soil. We aim for a simple, not numerically intensive model which could be used either as forward model in a Bayesian estimation scheme, or else as a preliminary means to identify key features of a concrete problem, for its further analysis by more sophisticated theoretical and numerical approaches.
\end{abstract}

\setboolean{displaycopyright}{true}

\begin{document}    
\maketitle

\section{Introduction} \label{sec:Introduction}

The scattering of electromagnetic waves in random dielectric media is a subject that continues to be challenging at present. This scenario can be found both at the microscopic scale where the scattering medium is illuminated by wavelengths in the visible or infrared spectrum and at regional scales where the earth's surface is monitored with orbital or airborne radars. Whatever the case, the typical lengths at which a medium can be considered homogeneous or have anisotropies are always defined with respect to the wavelength of the incident signal. Moreover, to extract useful information from the data it is necessary to have a theoretical model connecting the measured quantities to the physical parameters of the target. 

Our goal in this paper is to develop such a theoretical model in order to be able to compute, from the assumed geometric and statistical properties of a vegetation layer, the intensity of backscattered radiation in all HH, VH, VV and HV channels. This means that the incident radiation belongs to the microwave spectrum and has a wavelength larger than the characteristic correlation lengths of the medium. Our model assumes this and so it would not be directly applicable in other wavelengths.

The two considerations that guide our work are that we want a coherent model, and a numerically non-intensive model which could be used as an input in a Bayesian estimation scheme \cite{lee1989bayesian,gregory2005bayesian,heard2021introduction}. For this reason we aim for an analytic model, instead of a model based on intensive numerical simulations. We include in this latter category approaches where the forward model is used to build a data cube which is then the input for the estimation scheme. For an extensive review of both analytical and numerical models see \cite{tsang2000scattering,tsang2004scattering,tsang2004scattering_num,long2015microwave}.

In the context of remote sensing, where the earth's surface is monitored at scales of the order of kilometers, the randomness of the scattering medium can be described mainly by three factors: the roughness of the surface, the stratification of the underlying medium, and the vegetation on the surface. Regarding the first two scattering mechanisms, there is an extensive literature describing how to  calculate the amplitude and phase of the scattered wave under different types of approaches \cite{tsang2004scattering, voronovich2013wave,pinel2013electromagnetic,ishimaru2017electromagnetic}. The different types of theories take into account the wavelength of the incident signal and the parameters indicating the randomness of the different types of scenarios, the most important being the roughness of the scattering surface or the inhomogeneities of the material medium. The third scattering mechanism turns out to be the most complicated to calculate analytically. The vegetation on the surface is usually modeled as a dielectric medium with random inhomogeneities or, alternatively as a collection of randomly oriented cylinders. The propagation and scattering of electromagnetic waves in such media is a problem in itself. To deal with these issues, two of the most commonly used approaches used in remote sensing are the Radiative Transfer Equation (RTE) \cite{ulaby1981microwave,shin1989radiative,tsang2000scattering,ishimaru2017electromagnetic} and the Distorted Born Approximation (DBA) \cite{zuniga1980depolarization,jin1985ladder,borgeaud1987theoretical,borgeaud1989theoretical}. A well-known result is that for both RTE and DBA cross-polarization effects in the backscattered signal are only obtained if the calculations are performed to second order \cite{tsang2000scattering,zuniga1980depolarization,long2015microwave}. 

Although the RTE method may in principle be simpler to implement than DBA-based models, the disadvantage of the former is that it is not coherent: it only yields the backscattered energy of the incident signal. Nowadays this is a limitation since in the last decades full polarimetric orbital missions have been developed, i.e. radars that can measure both the amplitude and the phase of the received signals. Furthermore, they can measure in the different modes of incident and received polarization (i.e. four channels: HH, VH, VV and HV). For example, the RADARSAT-2 mission operates in C-band ($f=5.4~\text{GHz}$) and measures backscattered power on all four channels \cite{morena2004introduction}. The SAOCOM mission operates in L-band ($f=1.23~\text{GHz}$) and measures amplitude and phase on all four channels \cite{saocom}. Therefore, in order to make full use of the information obtained by this type of mission, it is necessary to have direct models that can account for all the observed parameters (namely, the covariance matrix $C$ or the coherence matrix $T$ \cite{lee2017polarimetric}). By making full use of the information obtained from polarimetric missions, the inference of biophysical parameters of interest, such as soil moisture, can be substantially improved. For example, in \cite{arellana2023}  the dielectric constant of non vegetated soils is estimated from the $T$ matrix using the Small Perturbation Method in layered media as a direct model \cite{tabatabaeenejad2006bistatic,sanamzadeh2017}. Following this line of work, the next step is to consider the vegetation on the surface. This leads to have a theoretical model that allows to calculate the amplitude and phase of the backscattered signal in such a scenario. This article is oriented towards this goal. 

Because of the same reasons that have guided our preference for an analytic model, namely, that its use in a Bayesian scheme involves running the model a large number of times, we have aimed for the simplest possible theoretical framework. We have opted for a model based on field theory techniques, namely we compute the scattered intensities and phases from a Schwinger-Dyson equation where the characteristics of the vegetation layer are introduced through its thickness and the two-point correlation in the dielectric constant fluctuations. The complexity of the vegetation layer makes an accurate modeling of these correlations elusive, and a truly realistic model would necessitate a large number of parameters, thereby detracting from predictive power. Our experience is that given the large uncertainties in relevant parameters a qualitatively correct ansatz for the two point correlations, of which we will give examples below, equations (\ref{casos}) and (\ref{cilindros}), is all that is required for a successful model. Also by reducing the statistical features of the dielectric constant fluctuations to their two point correlations we are assuming these fluctuations are Gaussian; for a more general treatment of non Gaussian fluctuations see \cite{franco2015wave,lamagna2017functional}.

Once the problem has been formulated as a set of Schwinger-Dyson equations, we shall once again choose simplicity by seeking a solution through a second order Born approximation \cite{zuniga1980depolarization,jin1985ladder,borgeaud1987theoretical,borgeaud1989theoretical}. This is the most straightforward way to find fully analytical expressions for the scattered fields, whereby we find the scattered energy by taking the second moments of the first and second order fields. Actually  at this order in powers of the dielectric constant fluctuation we may find interference terms between the zeroth and fourth, and first and third, order solutions, but these are known to be negligible \cite{zuniga1980depolarization}. Given the order of magnitude of the scattering amplitudes, computing higher orders of the perturbative expansion is unwarranted.

Still the Born approximation yields the desired intensities as multiple integrals over the vegetation layer. For a simple enough form of the two-point fluctuation correlation these integrals are elementary but their numerical evaluation is nontrivial, not least because the integrands are strongly oscillatory. However, this same feature makes a stationary phase approximation accurate and much simpler to implement. As a check, we recover a factor two enhancement of the backscattered flux, as expected for coherent scattering \cite{jin1985ladder,van1995self,knothe2013flux}. 

In summary, we present in this contribution a fully coherent, analytic model of the backscattering intensity in all HH, HV, VH and VV channels, which is simple enough to allow repeated evaluation as a forward model in a  Bayesian inference scheme. This set up may be used to obtain relevant parameters such as soil humidity, or simply as a fast way to identify the best model to be employed as input in more sophisticated approaches \cite{kim2013models,huang2021band,tsang2022theory,jeong2023full,jeong2023wave}, among which we include the consideration of non Gaussian statistical fluctuations. Moreover, the model only yields the volume scattering from the vegetation layer, to which we must add the surface scattering. Given the intensities involved, these two effects simply add and no cross correlation needs to be considered.

This article is organized as follows. In Section \ref{Sec:Preliminaries} we establish the system of equations that must be solved to find the amplitude and phase of the wave scattered by a stratified and inhomogeneous medium. Then, in Section \ref{Sec:Pertur_Exp} we develop the perturbative scheme that we will use to solve the equations given above. Section \ref{Sec:TE} and Section \ref{Sec:TM} are devoted to compute up to second order the scattering amplitude when the incident field corresponds to Transverse Electric (TE) or Magnetic (TM) case, respectively. In Section \ref{Sec:mean_values} we give the mean values of the scattered energy at each order in perturbations and we discuss how to extend these results to compute polarimetric issues. In Section \ref{Sec:results} we show the main results obtained from the proposed model. Finally, Section \ref{sec:Conclusions} contains our conclusions and proposals for future developments arising from what has been show here.

\section{Preliminaries} \label{Sec:Preliminaries}

\begin{figure}[ht]
    \centering
    \includegraphics[width = 0.45\textwidth]{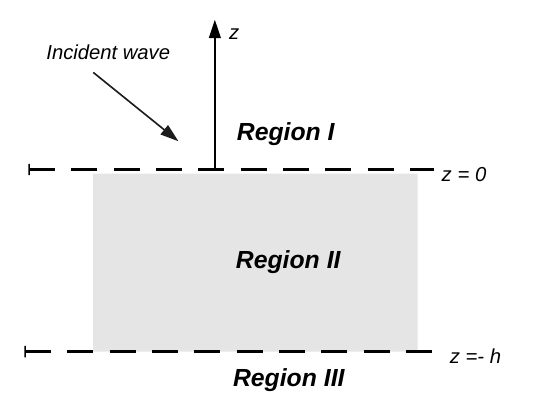}
    \caption{Geometry of the problem: the dielectric permittivity is piece-wise constant, with a discontinuity at $z=0$ and at $z=-h$. }
    \label{fig:geometry}
\end{figure}

Our first objective is to establish the basic scheme that we will use later to compute, through a perturbative scheme, the fields scattered by a stratified non-homogeneous target as we show in Figure \ref{fig:geometry}.

In this configuration there are three regions of interest: region I, from where the incident wave comes from; region II with flat interfaces at $z=0$ (with region I) and at $z=-h$ with region III; and region III, a semi-infinite medium at $z=-h$. Regions I and III are homogeneous with dielectric permittivity $\epsilon_{0I}$ and $\epsilon_{0III}$, respectively; the region II has a non-homogeneous permittivity that can be decomposed as its mean value plus fluctuation, i.e., $\epsilon_{II} = \epsilon_{0II} + \varepsilon_{II}(\mathbf{r})$.

\subsection{Problem Definition}\label{SubSec:Maxwell}

Our starting point is the Maxwell equations without sources in each region 
\be 
\mathbf{\Delta}\mathbf{E}-\nabla\left(\div\mathbf{E} \right) +\epsilon(\mathbf{r})\,\left(\frac{\omega}{c}\right)^2\mathbf{E}=0
\te 
where the dielectric permittivity $\epsilon(\mathbf{r}) = \epsilon(\vecx,z)$ indicates a piece-wise medium. We look for solutions of the form 
\be \label{Fourier_E}
\mathbf{E}(\mathbf{r}) = \int \frac{d^2p}{(2\pi)^2}\, e^{\imath \vecp\cdot\vecx}\,\mathbf{E}_{\vecp}(z)
\te 
being $\vecx=\left(x,y\right)$ and $\vecp=\left(p_x,p_y\right)$. Also we call $\mathbf{E}_{\vecp}=\left(\vec{E}_{\vecp},E_{z\,\vecp}\right)$ where further $\vec{E}_{\vec p}=\left(E_{x\,\vec{p}},E_{y\,\vec{p}}\right)$. The transverse component $\vec{E}_{\vecp}$ can be  decomposed into $H$ polarized modes where $\vec{E}_{H\vec{p}}=\left(\hat{z}\times\hat{p}\right)\varphi\left(z\right)$ and $V$ polarized case $\vec{E}_{V\vec{p}}=\hat{p}\,\psi\left(z\right)$, where $\hat{p}=\vec{p}/p$.

For the $H$ modes we have $\vec{p}\cdot\vec{E}_{H\vec{p}}=0$, and then
\be
\frac {\partial^2}{\partial z^2}\varphi +P^2\varphi =0
\label{KG}
\te
where $P^2=\epsilon(z) \left(\frac{\omega}{c}\right)^2-\vecp^2$ is the vertical component of the wave number in each space region. 

Similarly, for $V$ modes it results in
\be \label{Ez}
E_z=\frac{\imath p}{P^2}\frac {\partial\psi}{\partial z}
\te
and 
\be
\frac {\partial^2\psi}{\partial z^2}+\frac {\partial}{\partial z}\frac{p^2}{P^2}\frac {\partial\psi}{\partial z} +\epsilon \left(\frac{\omega}{c}\right)^2\psi=0
\label{KGV}
\te
where $\imath = \sqrt{-1}$ is the imaginary unit.

Observe that $E_{z}$ is defined from $\psi'$ evaluated at the same height.

Along with the dynamic equations for each mode we must give the matching conditions at the interfaces where the dielectric permittivity is not continuous. For the $H$ case we ask that $\varphi$ and $\varphi'$ are continuous over each interface; for the $V$ mode we require that $\psi$ and $\epsilon E_z$ are continuous. 

\subsection{Coupled equations and cross-pol effect}\label{SubSec:coupled_eqs}

We now consider the case where in each region of space the dielectric permittivity is written as $\epsilon = \epsilon_0\left(z\right)+\varepsilon\left(\vec{x},z\right)$, with $\epsilon_0\left(z\right)$ piece-wise constant. In a physical application there will be a finite illuminated area and there is no loss of generality in assuming that the dielectric constant fluctuations fall off smoothly outside this area. Under this assumption these fluctuations allow a two-dimensional Fourier transform \cite{ulmer2022monte}
\bea \label{def_ep}
&& \varepsilon_{\vecp}(z) = \int \frac{d^2x}{(2\pi)^2}\,e^{\imath \vecp\cdot\vecx}\,\varepsilon(\vecx,z) \nn
&& \varepsilon^*_{\vecp}(z) = \varepsilon_{-\vecp}(z) \,\,,\,\, \langle \varepsilon_{\vecp}(z) \rangle = 0
\tea 

The Fourier amplitudes are a Gaussian process with zero mean, and we will assume the correlation $ \langle \varepsilon(\mathbf{r})\,\varepsilon^*(\mathbf{r}')\rangle$ is translation invariant when its arguments are deeply within the illuminated area, so we may write
\bea \label{def_ep_corr}
&& \langle \varepsilon(\mathbf{r})\,\varepsilon^*(\mathbf{r}')\rangle = C(\mathbf{r}-\mathbf{r}') \nn 
&& \langle \varepsilon_{\vecp}(z)\,\varepsilon^*_{\vec{q}}(z')\rangle = \delta(\vecp-\vec{q})\,C_{\vecp}(z-z') 
\tea
where $C_{\vecp}(z)$ is the 2D Fourier transform of correlation function $C(\mathbf{r}-\mathbf{r}')$. Again  we decompose the Fourier expansion of the fields in terms of $H$ and $V$ modes, $\vec{E}_{\vec{p}}=\left(\hat{z}\times\hat{p}\right)\varphi_{\vec{p}}+\hat{p}\,\psi_{\vec{p}}$ and write these modes in terms of the orthogonal component of the electric field, so 
\begin{subequations}
 \be 
 \varphi_{\vec{p}} = \hat{z}\cdot\left(\hat{p}\times\vec E_{\vec p}\right)=-\hat{p}_y\,E_{x\,\vec{p}}+\hat{p}_x\,E_{y\,\vec{p}}
 \te 
 \be 
 \psi_{\vec{p}} = \hat{p}\cdot\vec E_{\vec p}=\hat{p}_x\,E_{x\,\vec{p}}+\hat{p}_y\,E_{y\,\vec{p}}
 \te 
\end{subequations}

Using the above relations, we get a Klein-Gordon type equation for each field
\begin{subequations}
\bea
&& \frac {\partial^2}{\partial z^2}\varphi_{X\vec{p}} +P^2\varphi_{X\vec{p}} \equiv -J_{XH\vec{p}}\nn
&& J_{XH\vec{p}} = {\left(\frac{\omega}{c}\right)^2}\int\frac{d^2q}{\left(2\pi\right)^2}\varepsilon_{\vec{p}-\hat{q}}(z)\nn
&&\left[\hat{p}\cdot\hat{q}\;\varphi_{X\vec{q}}+\hat{z}\cdot\left(\hat{p}\times\hat{q}\right)\psi_{X\vec{q}}\right]
\label{KG_H}
\tea 
and
\bea
&& \frac {\partial^2}{\partial z^2}\psi_{X\vec{p}} + \frac {\partial}{\partial z}\frac{p^2}{P^2}\frac {\partial}{\partial z}\psi_{X\vec{p}}+\epsilon_0\left(\frac{\omega}{c}\right)^2\psi_{X\vec{p}}\nn
&& \equiv -J_{XV\vec{p}} -\frac {\partial}{\partial z}J'_{XV\vec{p}}\nn 
&&J_{XV\vec{p}}  = {\left(\frac{\omega}{c}\right)^2}\int\frac{d^2q}{\left(2\pi\right)^2}\varepsilon_{\vec{p}-\vec{q}}(z)\nn
&&\left[\hat{z}\cdot\left(\hat{p}\times\hat{q}\right)\varphi_{X\vec{q}}-\hat{p}\cdot\hat{q}\;\psi_{X\vec{q}}\right] \nn
&&J'_{XV\vec{p}}= \imath p\left(\frac{\omega}{c}\right)^2\frac{1}{P^2}\int\frac{d^2q}{\left(2\pi\right)^2}\varepsilon_{\vec{p}-\vec{q}}(z)\,E_{z\,\vec{q}}
\label{KG_V}
\tea
\end{subequations}

Here the subscript $X$ indicates incident field polarization. Then
equations (\ref{KG_H}) and (\ref{KG_V}) show that it is possible to change the polarization state of the scattered field, as we shall show later, though not in backscattering condition to first order.
\section{Perturbative expansion} \label{Sec:Pertur_Exp}

As we discussed above, we consider the case where we have three regions. In region $I$, $z>0$, $\epsilon_0=\epsilon_{0I}$. In region $II$, $0>z>-h$, $\epsilon_0=\epsilon_{0II}$ . In region $III$, $-h>z$, $\epsilon_0=\epsilon_{0III}$ . The dielectric constant is not fluctuating in regions $I$ and $III$, and shows fluctuations $\varepsilon$ in region $II$. We shall seek an expansion in powers of $\varepsilon$. To this effect we write 
\bea
\varphi&=&\varphi^{\left(0\right)}+\varphi^{\left(s\right)}\nn
\psi&=&\psi^{\left(0\right)}+\psi^{\left(s\right)}\nn
E_z&=&E_z^{\left(0\right)}+E_z^{\left(s\right)}
\tea

The zero superscript fields are solutions to the homogeneous problem where a plane wave $\mathbf{E}=\mathbf{E}_{\mathbf{k}}e^{i\mathbf{k}\cdot\mathbf{r}}$ is incident from region I with wave number $\mathbf{k} = (\vec{k},-KI)$.  
Therefore the magnetic field (recall that $\left|\mathbf{k}\right|=\omega/c$) is $\mathbf{B}=\hat{\mathbf{k}}\times\mathbf{E}$ and the Poynting vector $\mathbf{S}=\hat{\mathbf{k}}\left|\mathbf{E}\right|^2$. In this condition, waves in region III are purely downgoing.

We have two situations, the horizontal polarization case (TE mode) where
\be
\varphi^{\left(0\right)}_{H\vec{p}}=\left(2\pi\right)^2\delta\left(\vec{p}-\vec{k}\right)\sqrt{2K_{I}}F^{\DWa}_{HK}\,\,,\,\,\psi^{\left(0\right)}_{H\vec{p}} = 0 \label{TE-incidence}
\te
and the vertical polarization case (TM mode) where we have
\be
\varphi^{\left(0\right)}_{V\vec{p}}=0 \,\,,\,\, \psi^{\left(0\right)}_{V\vec{p}}=\left(2\pi\right)^2\delta\left(\vec{p}-\vec{k}\right) \sqrt{2K_{I}}F^{\DWa}_{VK}
\label{TM-incidence}
\te
The functions $F^\DWa_{XK}$ are defined in equations (\ref{F_up}) and (\ref{F_down}) and they represent just a compact form to write the waves propagating into the layered media in the homogeneous case. In both modes $E^{\left(0\right)}_{zV\vec{p}}$ is obtained from equation (\ref{Ez}). 

For either $X=H,V$, the scattered fields $\varphi^{\left(s\right)}_X$ $\psi^{\left(s\right)}_X$ obey equations of the form of (\ref{KG_H}) or (\ref{KG_V}). Also, the vertical component of the electric field satisfies,
\bea
&& P^2E^{\left(s\right)}_{z\,X\vec{p}}+\left(\frac{\omega}{c}\right)^2\int\frac{d^2q}{\left(2\pi\right)^2}\varepsilon_{\vec{p}-\vec{q}}(z)\,E^{\left(s\right)}_{z\,X\vec{q}}=\nn
&&\imath p\frac d{dz}\psi^{\left(s\right)}_{X\vec{p}}-\left(\frac{\omega}{c}\right)^2\int\frac{d^2q}{\left(2\pi\right)^2} \varepsilon_{\vec{p}-\vec{q}}(z)\,E^{\left(0\right)}_{z\,X\vec{q}}
\label{KG_Z}
\tea 

The sources $J_{XH\vec{p}}$ and $J_{XV\vec{p}}$ in the right-hand sides of equations (\ref{KG_H}) and (\ref{KG_V}) are non vanishing in region $II$ only. The solutions can be expressed in terms of the Green functions of the homogeneous problem,
\be
G_{\left(H,V\right) \vec{p}}\left(z,z'\right)=\imath\,F^{\UPa}_{\left(H,V\right)p}\left(z_>\right)F^{\DWa}_{\left(H,V\right)p}\left(z_<\right)
\te

The $\UPa$ (\textit{up}) or $\DWa$ (\textit{down}) modes are purely outgoing solutions normalized to unit Wronskian. For $H$ and $V$ modes the Wronskian of two solutions is
\bea
&& W_{H1,2}=\left(-i\right)\left[\varphi_2\varphi'_1-\varphi_1\varphi'_2\right] \\
&& W_{V1,2}=\left(-i\right)\left[1+\frac{p^2}{P^2}\right]\left\{\psi'_1\psi_2-\psi_1\psi'_2\right\} 
\tea

For both modes ($m = H\,\text{or}\,V$) the outgoing waves, in each region of the space, can be written as follows,
\be \label{F_up}
\begin{cases}
F^{\UPa}_{m\,PI} = f_m(P_I)\,e^{\imath iP_{I}z} \\
F^{\UPa}_{m\,PII} = f_m(P_{II})\,\left[r^{\UPa}_{m}\,e^{\imath P_{II}z}+t^{\UPa}_{m}\,e^{-\imath P_{II}z}\right] \\ 
F^{\UPa}_{m\,PIII} = f_m(P_{III})\,\left[\frac{1}{t^{\DWa}_{m}}\,e^{\imath P_{III}z}+t^{\UPa}_{m}\,e^{-\imath P_{III}z}\right]
\end{cases}
\te

\be \label{F_down}
\begin{cases}
F^{\DWa}_{m\,PI} = f_m(P_I)\,\left[r^{\DWa}_{m}\,e^{\imath P_{I}z}+e^{-\imath P_{I}z}\right] \\
F^{\DWa}_{m\,P_{II}} = f_m(P_{II})\,\left[r^{\DWa}_{m}\,e^{\imath P_{II}z}+t^{\DWa}_{m}\,e^{-\imath P_{II}z}\right] \\
F^{\DWa}_{m\,P_{III}}=  f_m(P_{III})\, t^{\DWa}_{m}\,e^{-\imath P_{III}z} 
\end{cases}
\te
where the coefficients $r$ and $t$ for each mode are the Fresnel coefficients of reflection and transmission, and $f_m(P_\alpha) = \left\lbrace 1/\sqrt{2P_{\alpha}}, \sqrt{P_{\alpha}/(2\epsilon_{\alpha}(\omega/c)^2)} \right\rbrace $ is a normalization factor enforcing the unit Wronskian condition. Note that the Green functions obey the reciprocity condition \cite{van1998reciprocity},
\be
G_{\left(H,V\right) \vec{p}}\left(z,z'\right)=G_{\left(H,V\right) \left(-\vec{p}\right)}\left(z',z\right)
\te
which, since by rotation symmetry $G_{\left(H,V\right) \vec{p}}\left(z,z'\right)=G_{\left(H,V\right) {p}}\left(z,z'\right)$, reduces to symmetry under the exchange of $z$ and $z'$. 

The solutions for the scattered fields in region $I$ are
\begin{subequations}
\be\label{vp_I}
\varphi^{\left(s\right)}_{X\vec{p}}\left(z\right)=-F^{\UPa}_{H\,PI}\left(z\right)\int_{-h}^0dz'\;F^{\DWa}_{H\,PII}\left(z'\right)J_{XH\vec{p}}\left(z'\right)
\te 
\bea \label{psi_I}
&&\psi^{\left(s\right)}_{X\vec{p}}\left(z\right)=-F^{\UPa}_{V\,PI}\left(z\right)\int_{-h}^0dz'\;F^{\DWa}_{V\,PII}\left(z'\right)J_{XV\vec{p}}\left(z'\right)\nn
&&+F^{\UPa}_{V\,PI}\left(z\right)\int_{-h}^0dz'\;\frac{\partial F^{\DWa}_{V\,PII}}{\partial z'}\left(z'\right)J'_{XV\vec{p}}\left(z'\right)
\tea
\end{subequations}

In the last term we have made an integration by parts. Since we are assuming there are no fluctuations at the boundary of region II the integrated term vanishes.

In region $II$ the fields result
\begin{subequations}
\bea \label{vp_II}
&&\varphi^{\left(s\right)}_{X\vec{p}}\left(z\right) =\nn
&& -F^{\UPa}_{H\,PII}\left(z\right)\int_{-h}^z\,dz'\;F^{\DWa}_{H\,PII}\left(z'\right)J_{XH\vec{p}}\left(z'\right)\,\Theta(z>z') \nn
&&-F^{\DWa}_{H\,PII}\left(z\right)\int_{z}^0dz'\;F^{\UPa}_{H\,PII}\left(z'\right)J_{XH\vec{p}}\left(z'\right)\,\Theta(z<z') \nn 
\tea    
\bea \label{psi_II}
&& \psi^{\left(s\right)}_{X\vec{p}}\left(z\right) = \nn 
&& -F^{\UPa}_{V\,PII}\left(z\right)\int_{-h}^z \, dz'\;F^{\DWa}_{V\,PII}\left(z'\right)J_{XV\vec{p}}\left(z'\right)\,\Theta(z>z')\nn
&&+F^{\UPa}_{V\,PII}\left(z\right)\int_{-h}^z \, dz'\;\frac{\partial F^{\DWa}_{V\,PII}}{\partial z'}J'_{XV\vec{p}}\left(z'\right)\,\Theta(z>z')\nn
&& -F^{\DWa}_{V\,PII}\left(z\right)\int_{-h}^0\,dz'\;F^{\UPa}_{V\,PII}\left(z'\right)J_{XV\vec{p}}\left(z'\right)\,\Theta(z<z') \nn 
&& +F^{\DWa}_{V\,PII}\left(z\right)\int_{-h}^0\,dz'\;\frac{\partial F^{\DWa}_{V\,PII}}{\partial z'}\left(z'\right)J'_{XV\vec{p}}\left(z'\right)\,\Theta(z<z') \nn 
\tea 
\end{subequations}
\section{TE incident mode} \label{Sec:TE}

To study the TE incident mode we use as source field equations (\ref{TE-incidence}). We will define two scattering amplitudes: $S_{HH}$ and $S_{VH}$ for the $H$-mode and $V$-mode scattered fields, respectively. Each of these amplitudes will be computed order by order in terms of the perturbations in the dielectric permittivity of region II as well as the lower order $\varphi$ and $\psi$ fields. Namely, the solution to first order will be proportional to the product between the zero-order fields and the dielectric fluctuations, the second order solution will be proportional to the first-order fields and the dielectric fluctuations, and so on.

In order to find the scattered field in $H$ mode at first order we use equation (\ref{vp_I}) evaluating the source with the field given by (\ref{TE-incidence}). Similarly, the $V$ mode is obtained using (\ref{psi_I}) with the incident field (\ref{TE-incidence}). 
The first-order results evaluated in region $II$, which are given by equations (\ref{vp_II}) and (\ref{psi_II}), will be used as sources to find the scattered fields to second order. 

To simplify the upcoming formulae, we introduce the following notations. An index 
$\mu=\{\UPa, \DWa\}(\Rightarrow\bar{\mu}=\{\DWa,\UPa\})$ indicates the propagation mode of waves in region II, $\Theta_{\UPa}(z_1,z_2) = \Theta(z_1-z_2)(\Rightarrow\Theta_{\DWa}(z_1,z_2) = \Theta(z_2-z_1))$ shows where the second scattering process occurs. Index $m=\{H, V\}$ indicates the polarization of waves in medium II.

The H-mode scattered field results in,
\bea \label{HH_O2}
&& \vpp^{(2)}(z) = -\sqrt{2K_I}\,\left(\frac{\omega}{c}\right)^4\,F^{\UPa}_{HPI}(z)\nn 
&& \int\,\frac{d^2q}{(2\pi)^2}\,\int^0_{-h}\,dz_1\,\int^{0}_{-h}\,dz_2\,\,\EPpq(z_1)\,\EPqki(z_2)\nn
&& F^{\DWa}_{HPII}(z_1)\,\left[\sum_{\mu\,m}\beta_m\,F^\mu_{mQII}(z_1)\right.\nn
&& \left. F^{\hat{\mu}}_{mQII}(z_2)\,\Theta_\mu(z_1,z_2)\right]\,F^{\DWa}_{HKII}(z_2)  
\tea
where we have defined the geometric coefficients $\beta_H = (\hat{p}\cdot\hat{q})\,(\hat{q}\cdot\hat{k})$ and 
$\beta_V = \hat{z}\cdot\left(\hat{p}\times\hat{q}\right)\,\hat{z}\cdot\left(\hat{q}\times\hat{k}\right)$.

Before to get the V-mode scattered field it it convenient to define a compact form to write the derivative of $F$ functions:
\be \label{tilde_F}
\frac{dF^\mu_{m Q}}{dz}(z) = \imath Q \tilde{F}^\mu_{m Q}(z)
\te 
where $\tilde{F}$ has the same expression that in (\ref{F_up}) or (\ref{F_down}) but with the only difference that the coefficients $t^\mu_m$ are multiplied by a factor $(-1)$. Then, we can write the V scattered mode as,
\bea
&& \psi^{(2)}_{\vecp}(z) =- \sqrt{2KI}\left(\frac{\omega}{c}\right)^4\,F^{\UPa}_{V PI}(z)\nn 
&& \int^0_{-h}\,dz_1\,\int^0_{-h}\,dz_2\,\int \frac{d^2q}{(2\pi)^2}\EPpq(z_1)\,\EPqki(z_2)\nn
&& F^{\DWa}_{V PII}(z_1)\left[\sum_{\mu\,m}\gamma_m F^\mu_{m QII}(z_1)\right. \nn 
&& \left. F^{\bar{\mu}}_{m QII}(z_2)\Theta_\mu(z_1,z_2)\right] F^D_{H KII}(z_2) \label{VH_O2}
\tea
where the index $m$ now takes three values $\left\lbrace H, V, \tilde{V}1 \right\rbrace$. The last value comes from the second line of (\ref{psi_I}) after integration by parts and it indicates that the two functions $F_V$ evaluated at $z_1$ has been derived and must be written through (\ref{tilde_F}). Here  The corresponding coefficients are $\gamma_H = \hat{z}\cdot\left(\hat{p}\times\hat{q}\right)\,(\hat{q}\cdot\hat{k})$, $\gamma_V =- \left(\hat{p}\cdot\hat{q}\right)\,\hat{z}\cdot(\hat{q}\times\hat{k})$ and $\gamma_{\tilde{V}_1} = p\,q\,(PII\,QII)^{-1}\,\hat{z}\cdot(\hat{q}\times\hat{k})$.

\section{TM incident mode} \label{Sec:TM}

To study the TM incident mode we shall proceed in the same way than we did in Section \ref{Sec:TE}. Now we use the field given by equations (\ref{TM-incidence}) as source. We will define two new scattering amplitudes: $S_{HV}$ and $S_{VV}$, which correspond to $H$ mode and $V$ mode scattered fields, respectively. 

We use the incident field  given by (\ref{TM-incidence}) to evaluate the currents. To obtain the scattering amplitude for the HV channel to first order we use (\ref{vp_I}).  We then use
the first-order fields as sources to write the propagating waves in region II with equations (\ref{vp_II}) and (\ref{psi_II}) and compute the currents. 

The H-mode scattered field results in,
\bea
&& \vpp^{(2)} = -\sqrt{2K_I}\,\left(\frac{\omega}{c}\right)^4\,F^{\UPa}_{HPI}(z)\nn 
&& \int\,\frac{d^2q}{(2\pi)^2}\,\int^0_{-h}\,dz_1\,\int^{0}_{-h}\,dz_2\,\EPpq(z_1)\,\EPqki(z_2)\nn
&& F^{\DWa}_{HPII}(z_1)\left[ \sum_{\mu\,m}\,\delta_m\,F^\mu_{mQII}(z_1)\right.\nn 
&& \left. F^{\hat{\mu}}_{mQII}(z_2)\,F^{\DWa}_{V KII}(z_2)\,\Theta_\mu(z_1,z_2) \right] \label{HV_O2}
\tea
with coefficients $\delta_H = (\hat{p}\cdot\hat{q})\,\hat{z}\cdot(\hat{q}\times\hat{k})$,
$\delta_V = \hat{z}\cdot(\hat{p}\times\hat{q})\,(\hat{q}\cdot\hat{k})$ and $\delta_{\tilde{V}2} = -\hat{z}\cdot\left(\hat{p}\times\hat{q}\right)\,q\,k(QII\,KII)^{-1}$.

For the V-mode scattered field we find,
\bea
&& \psi^{(2)}_{\vec{p}} (z)= \sqrt{2K_I}\,\left(\frac{\omega}{c}\right)^4\,F^{\UPa}_{V PI}(z)\nn 
&& \int\,\frac{d^2q}{(2\pi)^2}\,\int^0_{-h}\,dz_1\,\int^0_{-h}\,dz_2\,\,\EPpq(z_1)\,\EPqki(z_2)\nn
&& \sum_{\mu\,m}\,\eta_m\,F^{\DWa}_{V PII}(z_1)\,F^\mu_{m QII}(z_1)\nn 
&& F^{\hat{\mu}}_{m QII}(z_2)\,F^{\DWa}_{V KII}(z_2)\,\Theta_\mu(z_1,z_2) \label{VV_O2}
\tea
Here the index $m$ takes five values: $\left\lbrace H, V, \tilde{V}_1, \tilde{V}_2, \tilde{V}_{12} \right\rbrace$; the last index means that the four functions $F$ with polarization $V$ have tilde and must be written as (\ref{tilde_F}). The geometric coefficients are:
$\eta_H = -\hat{z}\cdot(\hat{p}\times\hat{q})\,\hat{z}\cdot(\hat{q}\times\hat{k})$, $\eta_V = (\hat{p}\cdot\hat{q})\,(\hat{q}\cdot\hat{k})$, $\eta_{\tilde{V}1}=-(\hat{p}\cdot\hat{q})\,p\,k(PII\,KII)^{-1}$, $\eta_{\tilde{V}2}= (\hat{q}\cdot\hat{k})\,p\,q(PII\,QII)^{-1}$ and $\eta_{\tilde{V}12}=-p\,q^2\,k(PII\,(QII)^2\,KII)^{-1}$.

\section{Mean values} \label{Sec:mean_values}

\subsection{First Order} \label{Sec:mean_values_O1}

The scattering amplitude for HH, VH and HV channels can be written as
\bea
&& S^{(1)}_{ab}(z) = \sqrt{2KI}\,\left(\frac{\omega}{c}\right)^2\,\tau_{ab}\,f_a(PII)\,f_b(KII)\,F^{\UPa}_{a\,PI}(z)\nn
&&\int^0_{-h}\,dz'\,\EPpk(z')\,\left[r^\DWa_a\,e^{\imath PII z'} + t^\DWa_a\,e^{-\imath PII z'} \right]\nn
&& \left[r_b\,e^{\imath KII z'} + t_b\,e^{-\imath KII z'} \right]
\tea 
where $a\,(b)$ is the polarization of the scattered (incident) wave, $f_a(PII)\,f_b(KII)$ is a factor coming from the product between the $F$ functions involved in the scattering amplitude and $r^\DWa$ ($t^\DWa$) is the reflection (transmission) coefficient of downward waves with polarization $a$ or $b$ evaluated at the $PII$ or $KII$ mode, respectively. From now on we will replace $\omega/c$ by $k_0$, the incident wave number in region I. The coefficient $\tau_{ab}$ is just a geometric factor that can be easily read from equations (\ref{KG_H}) and (\ref{KG_V}); for the co-polarized case it is $(\hat{p}\cdot\hat{k})$ while for the cross cases results $\hat{z}\cdot(\hat{p}\times\hat{k})$. This shows that, as we mentioned above, no cross polarization occurs at first order in backscattering condition.

The mean scattered fields vanish in virtue of (\ref{def_ep}). For the scattered mean energy we must compute:
\bea \label{mean_energy_O1}
&& \langle \big\vert S^{(1)}_{ab}(z) \big\vert^2 \rangle = \delta_{\vec{k}}(0)\,2 KI \, k^4_0\,\vert \tau_{ab}\vert^2\,\vert F^{\UPa}_{a PI} \vert^2 \nn
&& \vert f_a(PII)\vert^2\,\vert f_b(KII)\vert^2\,\int^0_{-h}\,dz_1\,\int^0_{-h}\,dz_2 \nn
&& C_{\mbp-\mbk}(z_1-z_2)\,\left[|r_a|^2\, e^{\imath (z_1-z_2) PII}  \right. \nn 
&& \left. + |t_a|^2\, e^{-\imath (z_1-z_2)PII} + 2  \Re\left\lbrace r_a\,t^*_a e^{\imath (z_1+z_2)PII} \right\rbrace \right] \nn
&& \left[|r_b|^2\, e^{\imath (z_1-z_2)KII} + |t_b|^2\, e^{-\imath (z_1-z_2)KII} \right. \nn 
&& \left. + 2  \Re\left\lbrace r_b\,t^*_b e^{\imath (z_1+z_2)KII} \right\rbrace\right]
\tea

Here $\delta_{\vec{k}}(0)$ indicates the area illuminated by the incident wave from region I. 

Finally, for VV case we must note from (\ref{KGV}) there is an extra term coming from the term proportional to $E_z$ given by (\ref{Ez}). Besides that, the mean energy can be written similarly to (\ref{mean_energy_O1}) with the difference that, instead of making the product of two functions $F$ evaluated at $z_1$ by two functions $F$ evaluated at $z_2$, we must make the product between $F_a\,F_b + \tilde{F}_a\,\tilde{F}_b$ at $z_1$ and the corresponding one evaluated at $z_2$. This will produce an expression analogous to (\ref{mean_energy_O1}) but with more terms in the double integral and without the global factor $\vert f_a\,f_b\vert^2$ but with terms dependent on $\vert f_a\,f_b\vert^2$, $\vert\tilde{f}_a\,\tilde{f}_b\vert^2$, $\left(f_{a}\,f_b\right)\,\left(\tilde{f}_a\cdot\tilde{f}_b\right)^*$ and its complex conjugate.

\subsection{Second Order} \label{Sec:mean_values_O2}

The four channels can be arranged as
\bea
&& S^{(2)}_{ab}(z) = -\sqrt{2K_I}\,k^4_0\,F^{\UPa}_{a\,PI}(z)\nn 
&&\int\,\frac{d^2q}{(2\pi)^2}\,\int^0_{-h}\,dz_1\,\int^0_{-h}\,dz_2\,\,\EPpq(z_1)\,\EPqki(z_2)\nn
&&\sum_{\mu,m}\,F^{\DWa}_{a\,PII}(z_1)\,\mathcal{H}^{\mu,m}_{\vecp\,\vecq}(z_1,z_2)\,F^{\DWa}_{b\,KII}(z_2)
\tea
where the function $\mathcal{H}^{\mu,m}_{\vecp\,\vecq}(z_1,z_2)$ contains all dependence on geometrical factors and on the products between $F$ and/or $\tilde{F}$ functions as well as the Heaviside functions for each propagation mode.

At second order, the scattering amplitudes depend on the product between permittivity fluctuations evaluated at different points of region II, representing multiple scattering processes. Therefore, the mean field in no longer zero, but the mean energy is non vanishing in the specular direction only.

To compute the average energy to second order we start from
\bea \label{mean_energy_wick}
&& \langle |S^{(2)}_{ab}(z)|^2 \rangle = 2K_I\,k^8_0\,|F^{\UPa}_{a\,PI}(z)|^2\nn 
&& \int\,\frac{d^2q}{(2\pi)^2}\int\frac{d^2q'}{(2\pi)^2}\int^0_{-h}dz_1\int^0_{-h} dz_2\int^0_{-h} dz'_1 \int^0_{-h}dz'_2\nn
&& \langle \EPpq(z_1)\,\EPqki(z_2)\, \varepsilon^*_{\vecp-\vecq'}(z'_1)\,\varepsilon^*_{\vecq'-\vec{k}}(z'_2)\rangle \nn
&& \left[\sum_{\mu,m}\,F^{\DWa}_{a\,PII}(z_1)\,\mathcal{H}^{\mu,m}_{\vecp\,\vecq}(z_1,z_2)\,F^{\DWa}_{b\,KII}(z_2)\right]\nn 
&&\left[\sum_{\mu',m'}\,F^{\DWa}_{a\,PII}(z'_1)\,\mathcal{H}^{\mu',m'}_{\vecp\,\vecq'}(z'_1,z'_2)\,F^{\DWa}_{b\,KII}(z'_2)\right]^*
\tea 


To develop this calculation we will use Wick's theorem: the expectation value of the product of four fluctuations in the dielectric constant of Region II is written as the sum of three terms. One of them will give the coherent contribution to the mean energy and will be in the specular direction with respect to the incident field. Consequently, the incoherent contribution to the mean scattered energy at second order can be written as the sum of two contributions. Compactly, this reads,
\be
\langle |S^{(2)}_{ab}(z)|^2 \rangle - \big\vert \langle S^{(2)}_{ab}(z) \rangle \big\vert^2 \equiv \mathcal{L}_{ab} + \mathcal{C}_{ab} \label{mean_energy_O2}
\te
where $\mathcal{L}_{ab}$ is known as the ladder term and $\mathcal{C}_{ab}$ as the cross term.





\begin{figure}[H]
\centering
\includegraphics[width = 0.45\textwidth]{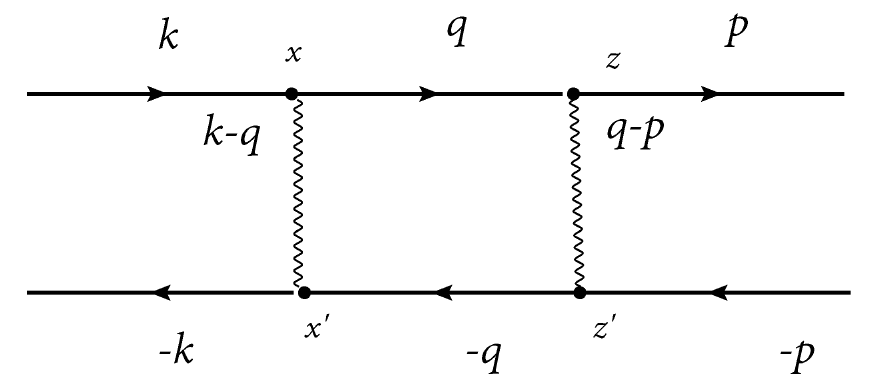}
\caption{Ladder diagram representing equation (\ref{lab}).}
\label{fig:ladder}
\end{figure}

In the above expression we defined the incoherent energy flux at second order as the sum of two contributions. Namely, when the auxiliary modes $\vecq$ and $\vec{q'}$ are equals we obtain the so-called Ladder term, which it is written as,
\bea 
&& \mathcal{L}_{ab} = \delta_{\vec{k}}(0)\, 
2K_I\,k^8_0 |F^{\UPa}_{a\,PI}(z)|^2\nn 
&&  \int\frac{d^2q}{(2\pi)^2}\int^0_{-h}dz_1\int^0_{-h} dz_2\int^0_{-h}dz'_1\int^0_{-h} dz'_2\nn
&& C_{\vecp-\vecq}(z_1-z'_1)\,C_{\vecq-\vec{k}}(z_2-z'_2)\nn
&&\sum_{\mu,m}\,\sum_{\mu',m'}\, \left[F^{\DWa}_{a\,PII}(z_1)\,\mathcal{H}^{\mu,m}_{\vecp\,\vecq}(z_1,z_2)\,F^{\DWa}_{b\,KII}(z_2)\right]\nn
&&\left[ F^{\DWa}_{a\,PII}(z'_1)\,\mathcal{H}^{\mu',m'}_{\vecp\,\vecq}(z'_1,z'_2)\,F^{\DWa}_{b\,KII}(z'_2)\right]^*
\label{lab}
\tea 
This scattering process is depicted in Figure \ref{fig:ladder}.

\begin{figure}[H]
\centering
\includegraphics[width = 0.45\textwidth]{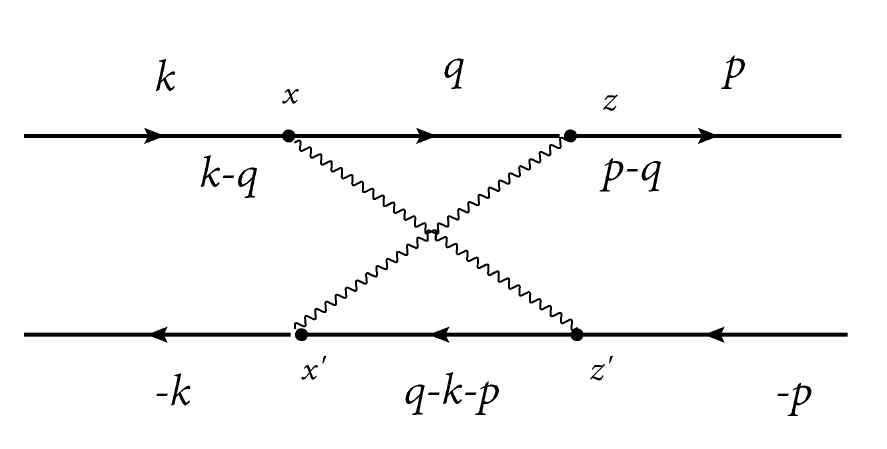}
\caption{Cross diagram representing to equation (\ref{cab}).}
\label{fig:cross}
\end{figure}

The second contribution is known as cross term and it is obtained when one of the auxiliary modes is evaluated at $\vec{u} = \vecp + \vec{k} - \vecq$, depicted in Figure \ref{fig:cross}
\bea 
&& \mathcal{C}_{ab} = \delta_{\vec{k}}(0)\, 
2K_I\,k^8_0 |F^{\UPa}_{a\,PI}(z)|^2\nn 
&&  \int\frac{d^2q}{(2\pi)^2}\int^0_{-h}dz_1\int^0_{-h} dz_2\int^0_{-h}dz'_1\int^0_{-h} dz'_2\nn
&& C_{\vecp-\vecq}(z_1-z'_2)\,C_{\vecq-\vec{k}}(z_2-z'_1) \nn
&& \sum_{\mu,m}\,\sum_{\mu',m'}\,\left[F^{\DWa}_{a\,PII}(z_1)\,\mathcal{H}^{\mu,m}_{\vecp\,\vecq}(z_1,z_2)\,F^{\DWa}_{b\,KII}(z_2)\right]\nn
&&\left[ F^{\DWa}_{a\,PII}(z'_1)\,\mathcal{H}^{\mu',m'}_{\vecp\,\vec{u}}(z'_1,z'_2)\,F^{\DWa}_{b\,KII}(z'_2)\right]^*
\label{cab}
\tea    

With the expressions (\ref{lab}) and (\ref{cab}) we complete, at least formally, the calculation of the mean energy scattered to the region I. This expression could be evaluated numerically, at a high computational cost. In the following, we shall present an approximation which is nevertheless accurate enough for comparison against field data.

To this end we will use a stationary phase approximation to the integrals over oscillatory functions, reducing the problem to a double integral over the intermediate impulse variable $\vecq$.

\subsubsection{Stationary phase approximation}

To get a compact expression for ladder a cross terms we introduce the following notation:
each function $F$ propagating with mode $\mu$ through region $II$, has polarization $m$ and depends of mode $Q$, is written as: 
\bea
&& F^\mu_{m\,Q}(z) = f_m(Q)\,\left[r^\mu_m(Q)\,e^{\imath Q z}+t^\mu_m(Q)\,e^{-\imath Q z}\right]  \nn 
&& F^\mu_{m\,Q}(z) = f_m(Q)\,\sum_{s=\pm 1}\,X^{\mu s}_{m\,Q}\,e^{\imath s Q z }
\label{Fmu_nocompacta}
\tea 
where $X^{\mu\,+1}_{m\,Q} = r^{\mu}_m(Q)$ and $X^{\mu\,-1}_{m\,Q} = t^{\mu}_m(Q)$. 

Then, in virtue of (\ref{tilde_F}), we get
\be
\tilde{F}^\mu_{m\,Q}(z) = f_m(Q)\,\sum_{s=\pm 1}\,s\,X^{\mu s}_{m\,Q}\,e^{\imath s Q z} \label{Fmu_compacta}
\te

The Ladder term given by equation (\ref{lab}) appears as a sum of many integrals, each integral being indexed by a different combination of the $s$ indexes introduced in eqs. (\ref{Fmu_nocompacta}) and (\ref{Fmu_compacta}), and the $\mu$ and $\mu'$ indexes introduced in (\ref{HV_O2}). All these integrals include oscillatory factors; the integrals where the phases of these factors are bounded away from zero are subdominant with respect to those where phases may reach zero. We keep only the main contributions to $\mathcal{L}_{ab}$ according to this criterion: for example, to eliminate the oscillation on $z_1$ we need $s_\mu = -s_1$ (then $PII=QII$) and to eliminate the oscillation on $z_2$ we need $s_2 = -s_{\bar{\mu}}$ (then $QII = KII$); analogous conditions arise on the primed variables. We use the above assumptions to simplify the sums over the indices $\left\lbrace s_\mu, s_{\bar{\mu}}, s'_\mu, s'_{\bar{\mu}} \right\rbrace $. After this we make the following change of variables:  $z_j = \xi_j + \frac{1}{2} \psi_j$, $z'_j = \xi_j - \frac{1}{2} \psi_j$. Then, we get

\bea
&& \mathcal{L}_{ab}  = \delta_{\vec{k}}(0)\,  2K_I\,k^8_0\,|F^{\UPa}_{a\,PI}(z)|^2 \nn
&& \int\,\frac{d^2q}{(2\pi)^2}\,\int^0_{-h}\,d\xi_1\,\int^h_{-h}\,d\psi_1\,\int\,d\xi_2\,\int\,d\psi_2\nn 
&& \sum_{s_1\,s_2}\,\sum_{\mu,m}\,\,\sum_{s'_1\,s'_2}\,\sum_{\mu',m'}\,\, C_{\vecp-\vecq}(\psi_1)\,C_{\vecq-\vec{k}}(\psi_2)\nn 
&& \Theta\left(s_1(\xi_1-\xi_2+\frac{1}{2}(\psi_1-\psi_2))\right)\nn 
&& \Theta\left(s'_1(\xi_1-\xi_2-\frac{1}{2}(\psi_1-\psi_2))\right)\nn
&& X^{\DWa\,s_1}_{a\,PII} \left(X^{\DWa\,s'_1}_{a\,PII}\right)^* |f_{a}(PII)|^2\nn 
&& \left[\beta_m\,\,|f_m(QII)|^2\,\,X^{\mu\,-s_1}_{m\,QII}\,X^{\bar{\mu}\,-s_2}_{m\,QII}\right] \nn
&& \left[\beta_{m'}\,\,|f_{m'}(QII)|^2\,\,X^{\mu'\,-s'_1}_{m'\,QII}\,X^{\bar{\mu}'\,-s'_2}_{m'\,QII}\right]^* \nn
&& X^{\DWa\,s_2}_{b\,KII} \left( X^{\DWa\,s'_2}_{b\,KII}\right)^*|f_{b}(KII)|^2 \nn
&& e^{\imath (PII-QII)(s_1-s'_1)\xi_1}\,e^{\imath \frac{1}{2}(PII-QII)(s_1+s'_1)\psi_1}\nn
&& e^{\imath(QII-\,KII)(s_2-s'_2)\xi_2}\,e^{\imath \frac{1}{2}(QII-\,KII)(s_2+s'_2)\psi_2}
\label{combinetas_O2}
\tea

Here it is worth noting that:

\begin{itemize}
    \item integration over $\xi_j$ takes its maximum value when $s_j = s'_j$;
    \item  when $\mu =\UPa \Rightarrow z_1 - z_2 > 0 \Rightarrow 2(\xi_1 - \xi_2) > \psi_2 - \psi_1$; and when $\mu' =\UPa\Rightarrow z'_1 - z'_2 > 0 \Rightarrow 2(\xi_1 - \xi_2) > \psi_1 - \psi_2$. A similar condition occurs when $\mu = \mu' = \DWa$. In both cases the condition $\psi_1 = \psi_2 = 0$ belongs to the integration domain and gives the maximum contribution to the integral since the oscillations on the $\psi$ variables cancel out;
    \item when $\mu = \UPa \Rightarrow z_1 - z_2 > 0 \Rightarrow 2(\xi_1 - \xi_2) > \psi_2 - \psi_1$; and when $\mu' =\DWa \Rightarrow z'_1 - z'_2 < 0 \Rightarrow 2(\xi_1 - \xi_2) < \psi_1 - \psi_2$. A similar condition occurs when $\mu = \DWa$ and $\mu' = \UPa$. In both cases the condition $\psi_1 = \psi_2 = 0$ it does not belong to the integration domain and therefore this combination of propagation modes will be subdominant in the total contribution to the ladder term.
\end{itemize}

Because of the above two points, from the equation (\ref{combinetas_O2}) we will only keep half of the terms: all those that are $\UPa\UPa$ and $\DWa\DWa$ adding up all possible combinations of polarization for propagating waves in medium II, which will give us a total of eight combinations.

Using that $\mu = \mu'$, $s_1 = s'_1$, $s_2 = s'_2$ and that one of the integrals over $\psi$ gives a factor $2h$, the ladder term reduces to
\bea
&& \mathcal{L}_{ab}  = \delta_{\vec{k}}(0)\,   4\,h\,K_I\,k^8_0\,|F^{\UPa}_{a\,PI}(z)|^2\,|f_{a}(PII)|^2 \nn 
&& |f_{b}(KII)|^2\, \int\,\frac{d^2q}{(2\pi)^2}\,\int^h_{-h}\,d\psi\,\int^0_{-h}\,d\xi_1\,\int^0_{-h}\,d\xi_2\,\nn 
&& \sum_{s_1\,s_2}\,\sum_{\mu,m,m'}\,\, C_{\vecp-\vecq}(\psi)\,C_{\vecq-\vec{k}}(\psi)\,\Theta\left(s_1(\xi_1-\xi_2)\right)\nn
&&|X^{\DWa\,s_1}_{a\,PII}|^2\,\left[\beta_m\,\,|f_m(QII)|^2\,\,X^{\mu\,-s_1}_{m\,QII}\,X^{\bar{\mu}\,-s_2}_{m\,QII}\right] \nn
&& \left[\beta_{m'}\,\,|f_{m'}(QII)|^2\,\,X^{\mu\,-s_1}_{m'\,QII}\,X^{\bar{\mu}\,-s_2}_{m'\,QII}\right]^* \,|X^{\DWa\,s_2}_{b\,KII}|^2\nn
&& e^{\imath s_1(PII-QII)\psi}\,e^{\imath s_2(QII-\,KII)\psi}
\tea


The integrals over $d\xi_1,d\xi_2$ give half the area $[-h,0]\times[-h,0]$ due to $\Theta\left(s_1(\xi_1-\xi_2)\right)$. The integral over $\psi$ is the product between three terms: the oscillating factor that depends on $s_1$, $s_2$ and $\psi$ and the two correlation functions evaluated at $\vecp-\vecq$ and $\vecq-\vec{k}$. From this integration we define,
\be \label{def_CI}
\mathcal{I}_{s_1\,s_2}(\Omega,\vecp-\vecq,\vecq-\vec{k}) = \int^h_{-h} d\psi\,e^{\imath \psi\,\Omega}\,C_{\vecp-\vecq}(\psi)\,C_{\vecq-\vec{k}}(\psi)
\te 
where $\Omega = s_1(PII-QII)+s_2(QII-KII)$ is the dominant mode in the stationary phase approximation. Below (see equations (\ref{acf_z_I}) and (\ref{acf_two_cyl})) we list the  correlation functions that we use to characterize the inhomogeneous medium. In most cases, the expression $\mathcal{I}_{s_1\,s_2}(\Omega)$ has an analytic form; when this is not possible we use an asymptotic approximation for it. After all, we arrive at a compact expression for the ladder term,
\bea
&&  \mathcal{L}_{ab} = \delta_{\vec{k}}(0)\, 
 2\,h^3\,K_I\,k^8_0\,|F^{\UPa}_{a\,PI}(z)|^2\nn 
&& |f_{a}(PII)|^2\,|f_{b}(KII)|^2\,\int\,\frac{d^2q}{(2\pi)^2}\nn 
&& \sum_{s_1,s_2}\,\sum_{\mu,m\,m'}\,\mathcal{I}_{s_1\,s_2}(\Omega,\vecp-\vecq,\vecq-\vec{k})\nn
&&|X^{\DWa\,s_1}_{a\,PII}|^2\,\left[\beta_m\,\,|f_m(QII)|^2\,\,X^{\mu\,-s_1}_{m\,QII}\,X^{\bar{\mu}\,-s_2}_{m\,QII}\right] \nn
&& \left[\beta_{m'}\,\,|f_{m'}(QII)|^2\,\,X^{\mu\,-s_1}_{m'\,QII}\,X^{\bar{\mu}\,-s_2}_{m'\,QII}\right]^* \,|X^{\DWa\,s_2}_{b\,KII}|^2\nn
&&  \label{Ladder_ab}
\tea


The integral over the intermediate impulse variable $\vecq$ is computed numerically. Here we see that the ladder term is real: when $m=m'$ all the coefficients $X$ appear as square module; when $m\neq m'$ we see that adding the complementary cases gives twice the real part of those terms.

The cross term is analyzed in a similar way. The details to note are that: one of the auxiliary modes is $\vec{u} = \vec{k}+\vec{p}-\vec{q}$ and that the vertical coordinates in the correlation functions are grouped as $z_1-z'_2$ and $z_2-z'_1$. Because of this, the propagating modes contributing to the stationary phase are $z_1-z'_2$. The other detail to note is that oscillations are suppressed only under the backscattering condition ($\vec{p} = - \vec{k}$), which implies $\vec{u} = -\vec{q}$ and $UII = QII$. Thus, the cross term takes the same expression that (\ref{Ladder_ab}) but with the integral defined in (\ref{def_CI}) evaluated at the mode $\Omega' = (s_1-s_2)(KII-QII)$. As can be easily checked, $\Omega'=\Omega$ in backscattering condition, i.e. $PII=KII$, whereby the cross and ladder terms are equal, leading to the so-called backscattering enhancement \cite{jin1985ladder}.

\subsection{Bistatic scattering coefficients}

So far we have presented the scattering amplitude at first and second order for both incident and scattered polarization. These amplitudes have been called $S_{HH}$ and $S_{VH}$ when we have a TE incident mode and the scattered field has  $H$ ($\varphi$) or $V$ ($\psi$) polarization; the other two combinations, $S_{HV}$ and $S_{VV}$, occur with a TM incident mode. The Poynting vector is $\mathbf{S} = \hat{\mathbf{k}}\,|\mathbf{E}|^2$. Now we introduce the bistatic scattering coefficient as,
\be \label{RCS}
\sigma_{ab}(\mathbf{p},\mathbf{k}) = \frac{1}{\mathcal{A}}\frac{\langle\mathbf{p}\cdot \mathbf{S}_a (\mathbf{p})\rangle_{\text{Incoh}}}{|\mathbf{E}_b(\mathbf{k})|^2} 
\te 

This magnitude represents the incoherent scattered energy with polarization $a$ in direction $\mathbf{p}$ normalized by unit area and incident energy where the incident wave has polarization $b$ and propagates in direction $\mathbf{k}$. 

As we discuss in Section \ref{Sec:mean_values}\ref{Sec:mean_values_O1} at first order the scattered energy is already incoherent. At second order we defined the incoherent scattered energy through the sum of the ladder and cross terms. Both first order mean energy, ladder and cross term are proportional to $\delta_{\vec{k}}(0)\, $ which represents the illuminated area by the incident wave. In addition to this, we have developed all our calculation using a unit incident electric field. Also, we are interested in analyzing the results in the backscattering condition ($\mathbf{p}=-\mathbf{k}$) where ladder and cross terms are equal. Therefore, the bistatic scattering coefficients will be written using expressions (\ref{mean_energy_O1}) and (\ref{Ladder_ab}). Both expressions depend on the 2D Fourier transform $C_{\vec{p}-\vec{k}}(z_1-z_2)$ of the correlation function  defined in (\ref{def_ep_corr}). 

To get a closed form to (\ref{mean_energy_O1}) and (\ref{Ladder_ab}) we choose an explicit expression for the correlation function $C_{\vec{p}}(z)$. To illustrate our results we will the same correlation function as in \cite{zuniga1980depolarization,zuniga1980active}:
\be \label{casos}
 C(\mathbf{r}-\mathbf{r}') = s
   \exp\{-\frac{|\vec{x}-\vec{x}'|^2}{l^2_r}\}\,\exp\{-\frac{|z-z'|}{l_z}\} 
\te

which correspond to vertical Exponential and radially Gaussian. In the above $s = \delta\,\Re\{\epsilon_{0II}\}^2$ represents the fluctuation intensity, $l_r$ ($l_z$) is the radial (vertical) correlation length. This function has the following 2D Fourier transform:
\be \label{acf_z_I}
 C_{\vecp}(z-z') = s\,
   \frac{\pi\,l^2_r}{(2\pi)^2}\,\exp\{-\frac{l^2_r}{4}\vecp^2\}\,\exp\{-\frac{|z-z'|}{l_z}\}
\te
From this expression it is straightforward to find $\mathcal{I}_{s_1,s_2}(\Omega)$.

Also we consider a correlation function for modeling overlapping randomly distributed vertical cylinders which have radius $R$ and height $H$ and low space density $\rho$. We assume the dielectric fluctuations at two points are correlated if they belong to the same cylinder, and uncorrelated otherwise. Therefore we need to compute the probability that two given points belong to a single cylinder; this probability is zero if the horizontal distance between the two points is larger than $2R$ or the vertical distance is larger than $H$. Otherwise this correlation can be written as follows,
\be \label{cilindros}
C(\mathbf{r}-\mathbf{r}') = \rho\,Z\,\Theta\left(Z\right)\,P(\chi)\,\Theta(1-\chi)
\te 
where $\chi = \frac{|r-r'|}{2R}$, $P(\chi) = \arccos(\chi)-\chi\sqrt{1-\chi^2}$ and $Z =1-\frac{|z-z'|}{2H}$. Its 2D Fourier transform is,
\be \label{acf_two_cyl}
C_{\vecp}(z-z') = \frac{2}{\pi}\,\rho\,Z\,\Theta(Z)\left[\frac{J_1(R|\vec{p}|)}{|\vec{p}|}\right]^2 
\te

At this point it is important to mention that what has been developed so far can easily be re-adapted to calculate both the coherence matrix $T$ or the covariance matrix $C$. Both matrices involve the calculation of mean values between the different $S_{qp}$ amplitudes.
For example, to calculate $\langle S_{HH}\,S^*_{VV} \rangle$ we can use the results given by equations (\ref{mean_energy_O1}) and (\ref{Ladder_ab}) by simply replacing one of the amplitudes by $S_{ab}$ by $S_{rs}$ and obtain the desired mean value. This change will only modify geometrical coefficients or reflection or transmission amplitudes, but will not affect in any way the development of the calculations made in the two previous subsections.

In the next section we will show the results obtained from the set up we have presented so far.

\section{Results} \label{Sec:results}

We illustrate our results through numerical simulations of different backscattering scenarios, where we either fix the dielectric and geometric parameters  and vary the incidence angle, or else we consider layered media with different geometric o dielectric parameters at a given incidence angle.

In order to validate our development we compare it with the results presented in \cite{zuniga1980active}. As we mentioned above, the copolarized cases (HH or VV) present nonzero backscattering amplitudes at first order in perturbations (just as in \cite{zuniga1980active}) meanwhile to get cross polarization effect it is needed to compute second order coefficients (similarly to \cite{zuniga1980depolarization}).

We show results to first order given by equation (\ref{mean_energy_O1}) for HH and VV channels. Here we compare the normalized backscattering coefficients (NBC) given by equation (\ref{RCS}) with the presented in Figure 4 of \cite{zuniga1980active}. Namely, we use 20 GHz as incident frequency, the three regions are characterized by $\epsilon_{0I} =1$, $\epsilon_{0II} =1.4 + \imath 0.02$, $\epsilon_{0III} =6 + \imath 0.6$, region II has thickness $h=3~\text{m}$; the in-homogeneous media has correlation lengths $l_r = 0.003~\text{m}$ and $l_z = 0.013~\text{m}$. In the same way as in \cite{zuniga1980active} we use a correlation function which is Gaussian laterally and exponential vertically.

\begin{figure}[!ht]
    \centering
    \includegraphics[width = 0.45\textwidth]{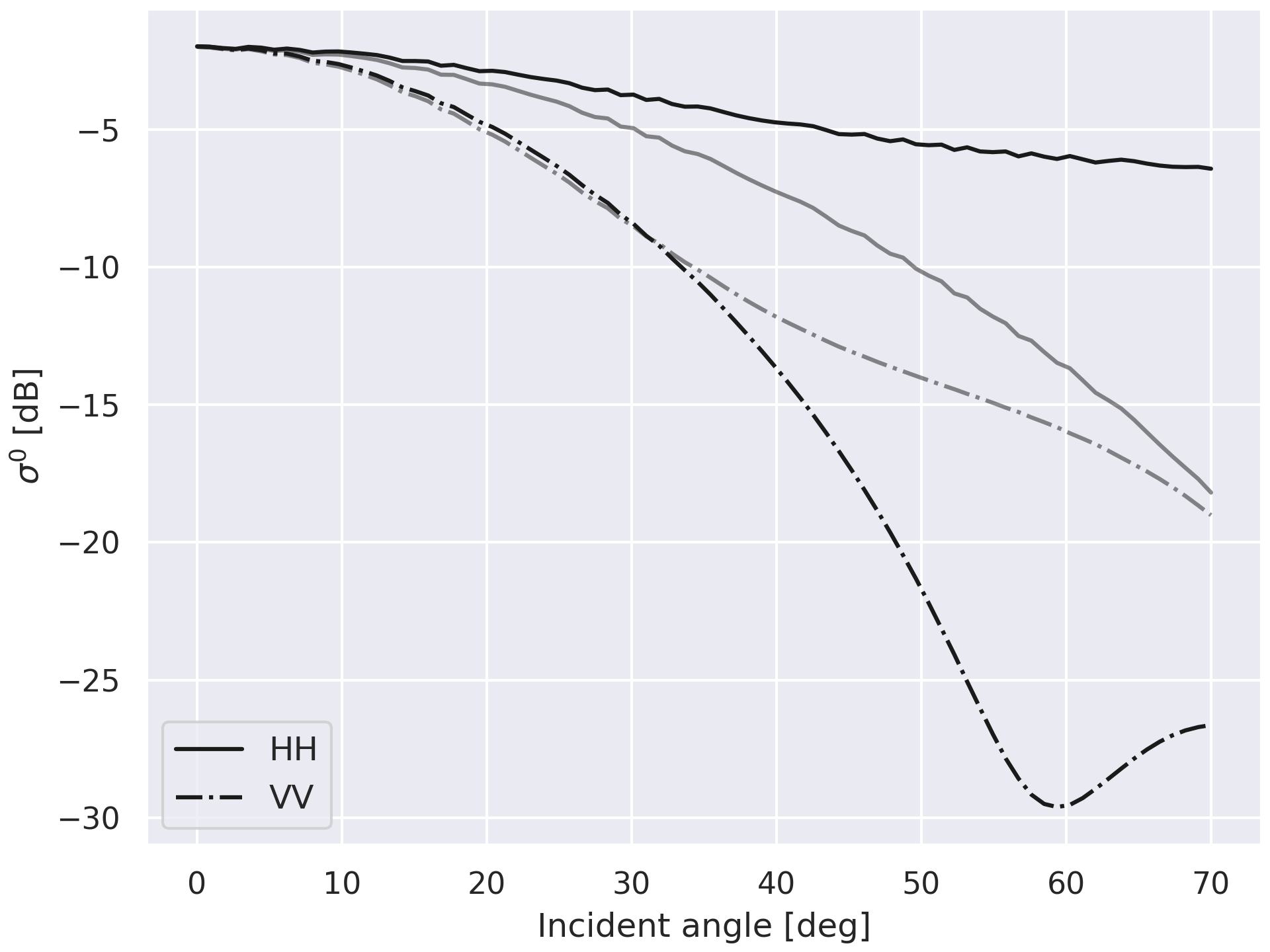}\\
    \caption{Normalized backscattering coefficient as a function of incident angle. Black (solid or dashed) lines correspond to our model, gray lines correspond to \cite{zuniga1980active}.}
    \label{fig:comparison_BS_theta_copol}
\end{figure}

In Figure \ref{fig:comparison_BS_theta_copol} we observe a similar trend for the vertically polarized case for small incident angles, meanwhile for the HH channel our model is less dependent on the angle of incidence than the one presented in \cite{zuniga1980active}. Our model also shows a strong suppression of the VV channel at incidence angles close to the Brewster angle between Regions II and III (i.e. $\theta_B \approx 64^\circ$ for $\epsilon_{0II} = 1.4$ and $\epsilon_{0III} = 6$).

\begin{figure}[!ht]
    \centering
    \includegraphics[width = 0.45\textwidth]{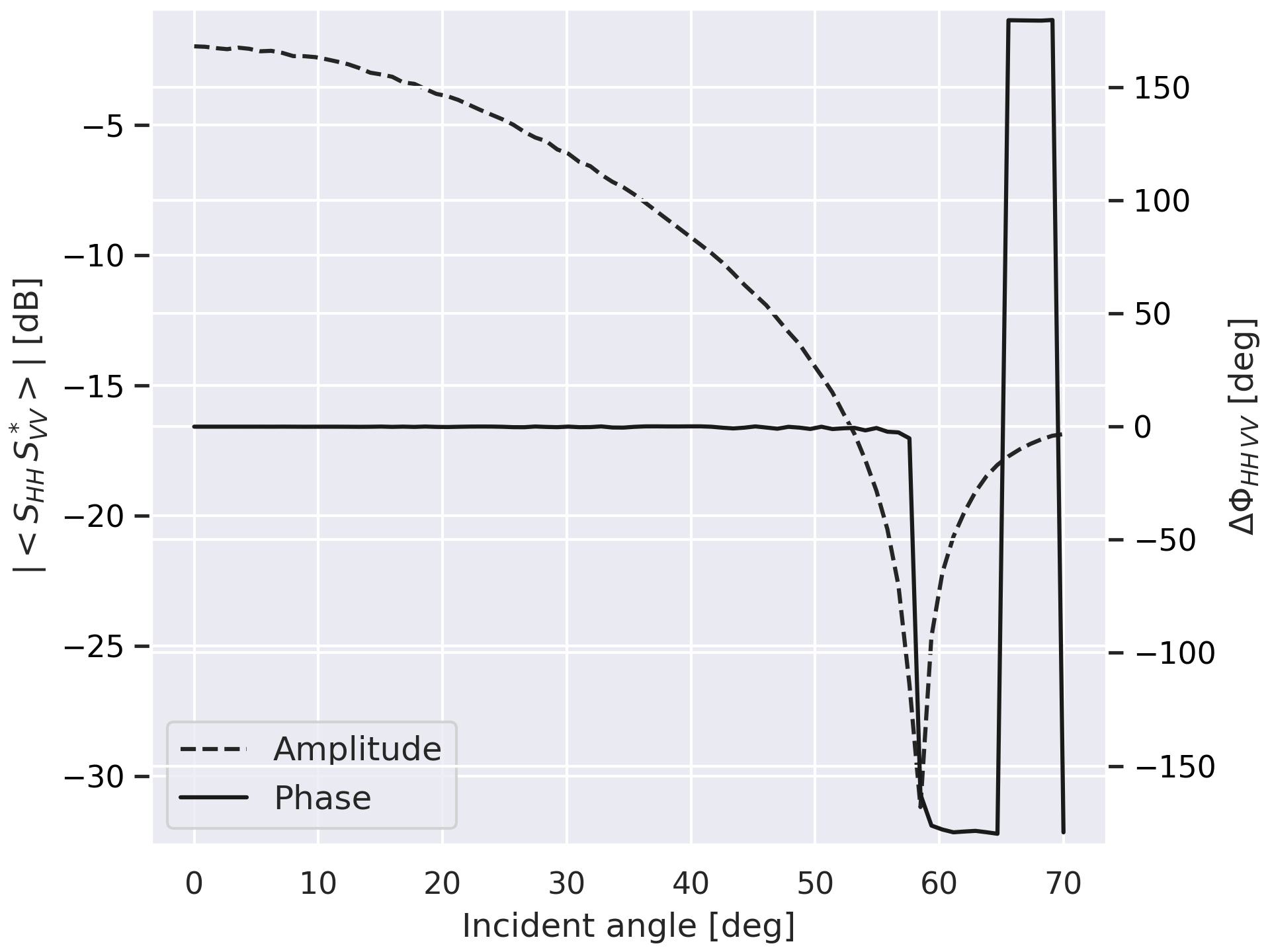}
    \caption{Phase difference between co-polarized channels using the same parameters as in Figure \ref{fig:comparison_BS_theta_copol}.}
    \label{fig:BS_theta_HH_VV}
\end{figure}

Using the same parameters as before we show in Figure \ref{fig:BS_theta_HH_VV} the module and phase of $\langle S_{HH}\,S^*_{VV}\rangle$. As expected, the amplitude is the same order of magnitude than in Figure \ref{fig:comparison_BS_theta_copol}. On the other hand, HH and VV are nearly in phase ($\Delta \phi_{HH VV} \approx 0$)  up to Brewster's angle between regions II and III and a change of $180^\circ$ appears in the relative co-polarized phase.

These results showed in Figures \ref{fig:comparison_BS_theta_copol} and \ref{fig:BS_theta_HH_VV} were computed with a desktop computer which has a processor Intel i7-9700 and four kernels. To compute the integration in (\ref{mean_energy_O1}) we use a Simpson quadrature scheme implemented in Python using Numpy and Scipy libraries \cite{numpy,2020SciPy-NMeth}. We have used 600 nodes to perform the numerical integration. Figures \ref{fig:comparison_BS_theta_copol} and \ref{fig:BS_theta_HH_VV} have 156 points in total and took 10.61 seconds to compute, therefore each point required 0.06 seconds. These times are adequate for use in a Bayesian inference scheme. The computation times for the Figures presented below were similar.

\begin{figure}[!ht]
    \centering
    \includegraphics[width = 0.45\textwidth]{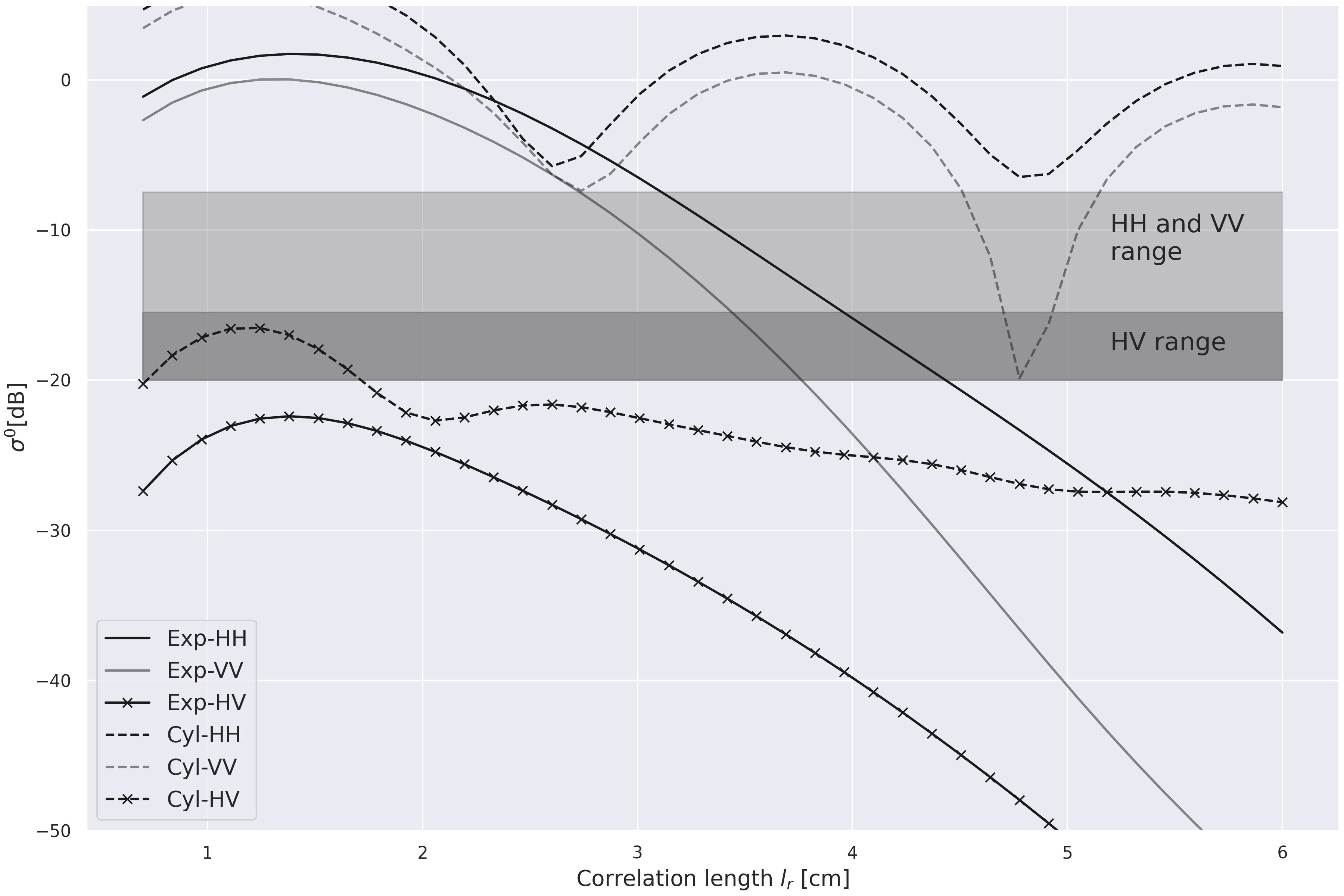}
    \caption{Backscattering coefficients for the three channels presented in \cite{morandeira2021}. Co-polarized range is defined between -15 and -7.5 dB; cross-polarized channel lies between -20 and -15 dB.}
    \label{fig:radarsat}
\end{figure}

To analyze if the development presented here is adequate for explain field data we use results from \cite{morandeira2021}. There, polarimetric RADARSAT-2 scenes were token over floodplain with incidence angle between 20 and 40 degrees with an incident frequency of 5.405 GHz. To characterize the background of region II we use a effective dielectric constant $\epsilon_{0II}=1.8$ and a layer thickness $h =1.65~\text{m}$, which is a mean of the height reported in \cite{morandeira2021}. The volumetric density of the cylinders was calculated from the field data reported in \cite{morandeira2021}, using the average radius and height of the cylinders and the number of cylinders per square meter. As for the dielectric constant $\epsilon_{0III}$ and correlation length $l_r$ and $l_z$ we varied the parameters taking into account the geometry to be explained. In Figure \ref{fig:radarsat} we present backscattering coefficients as a function of the radial correlation length $l_r$ keeping fixed $l_z = 2~\text{cm}$ and $\epsilon_{0III} = 20$.

In Figure \ref{fig:radarsat} solid lines correspond to the Exponential correlation and dashed lines to the overlapping cylinders. Black and grey lines correspond to HH and VV, respectively; the lines with a cross represent cross-polarized channels. From the above Figure it can be seen that the correlation between cylinders generates a higher depolarization, reaching almost the observed levels. Although it is not possible to generate $\sigma^0$ values so that the three channels are simultaneously within the corresponding range, this could be achieved by varying the model parameters that were left fixed. However, these same parameters can explain RADARSAT-2 observations presented in \cite{morandeira2016mapping}, where vegetation type similar to \cite{morandeira2021} is studied.

An exploratory analysis of the response of the model with respect to all its parameters is left for future work.

\section{Conclusions} \label{sec:Conclusions}

In this paper we have presented a model of scattering of microwave radiation by a random layer which is derived from first principles through a series of explicit approximations which are easily seen to hold for the case of microwave scattering by vegetation over bare soil. Our model is coherent, allowing the calculation of intensities and relative phases among all four channels HH, VV, HV and VH. Moreover, it is designed not to be numerically intensive, in view of its incorporation into inference schemes which demand running the core model a large number of times. We believe all these characteristics are necessary to make full use of existing observation capabilities, and in this sense this model opens up a number of applications to be presented in future contributions.

We have checked our model by computing the scattering intensities for two simple functional forms of the dielectric constant fluctuations correlation function, \ref{casos} and \ref{cilindros}. The results agree well with expectations from actual observations and show overall a common pattern. However there is a clear intensity gap between the results for the correlation function of overlapping cylinders \ref{cilindros} and the other ones \ref{casos}, and it would be an easy exercise to imagine further analytical expressions for the correlation function giving intermediate results between these extremes. This shows that a proper choice of the two-point correlation of the dielectric constant fluctuation is the single most important factor which determines the accuracy of the model, with an increased complexity in modeling (such as computing further orders in perturbation theory, including non Gaussian fluctuations, using a numerically intensive approach or replacing the analytical model by a numerical solution of Maxwell's equations from scratch) bringing only incremental gains.

Moreover, while in this paper we have focused on the backscattered intensities, the model yields full information about the scattered field phase and thus it can also be used in polarimetric analysis, which we will discuss in a separate contribution.

\section{Founding Information}

Work supported in part by CONICET, Universidad de Buenos
Aires UBACYT 20020220300204BA, CONICET PIP 11220210100595CO, ANPCYT PICT 2018-03684 and CONICET PICT 2020-01830.

\bibliography{biblio.bib}

\end{document}